\documentclass[aps,twocolumn,superscriptaddress,longbibliography]{revtex4-2}

\usepackage{graphicx}
\usepackage{CJK}
\usepackage{dcolumn}
\usepackage{bm}
\usepackage[colorlinks=true,bookmarks=false,citecolor=blue,linkcolor=blue,hyperfootnotes=true,urlcolor=blue]{hyperref}
\usepackage{color}
\usepackage{comment}
\usepackage{sidecap}
\usepackage{nicefrac}
\usepackage[acronym]{glossaries}

\begin{document}

\title{Visualization of Single Spin-chain Propagation using Bragg Coherent Diffraction Imaging}


\author{Xiaoqian M. Chen}
\affiliation{National Synchrotron Light Source II, Brookhaven National Laboratory, Upton, New York 11973, USA}


\author{Justin S. Woods}
\affiliation{Materials Science Division, Argonne National Laboratory, Lemont, IL 60439, USA.}

\author{Ulrich Welp}
\affiliation{Materials Science Division, Argonne National Laboratory, Lemont, IL 60439, USA.}

\author{Wai-kwong Kwok}
\affiliation{Materials Science Division, Argonne National Laboratory, Lemont, IL 60439, USA.}

\author{Ralu Divan}
\affiliation{Center for Nanoscale Materials, Argonne National Laboratory, Lemont, IL 60439, USA}

\author{David A. Czaplewski}
\affiliation{Center for Nanoscale Materials, Argonne National Laboratory, Lemont, IL 60439, USA}

\author{Scott Dhuey}
\affiliation{Molecular Foundry, Lawrence Berkeley National Laboratory, Berkeley, CA 94720, USA}

\author{Sujoy Roy}
\affiliation{Advanced Light Source, Lawrence Berkeley National Laboratory, Berkeley, CA 94720, USA}

\author{Rajesh V. Chopdekar}
\affiliation{Advanced Light Source, Lawrence Berkeley National Laboratory, Berkeley, CA 94720, USA}

\author{Lance E. De Long}
\affiliation{Dept. of Physics and Astronomy, University of Kentucky, Lexington, KY 40502, USA}

\author{Andi~Barbour}
\affiliation{National Synchrotron Light Source II, Brookhaven National Laboratory, Upton, New York 11973, USA}

\author{Claudio~Mazzoli}
\affiliation{National Synchrotron Light Source II, Brookhaven National Laboratory, Upton, New York 11973, USA}


\author{Ian Robinson}
\email{i.robinson@ucl.ac.uk}
\affiliation{Condensed Matter Physics and Materials Science Department, Brookhaven National Laboratory, Upton, New York 11973, USA}
\affiliation{London Centre for Nanotechnology, University College, Gower St., London, WC1E 6BT, UK}

\author{J.~Todd Hastings}
\email{todd.hastings@uky.edu}
\affiliation{Electrical and Computer Engineering, University of Kentucky, Lexington, KY 40502, USA}
\affiliation{Dept. of Physics and Astronomy, University of Kentucky, Lexington, KY 40502, USA}

\def\mathbi#1{\ensuremath{\textbf{\em #1}}}
\def\Q{\ensuremath{\mathbi{Q}}}
\def\q{\ensuremath{\mathbi{q}}}

\def\LBCO{La$_{1.875}$Ba$_{0.125}$CuO$_4$}

\newcommand{\angstrom}{\mbox{\normalfont\AA}}
\date{\today}

\begin{abstract}
Domain wall dynamics and their manipulation in finite-size systems serve as a promising platform for beyond Moore's law applications. Yet, the imaging techniques to track their motion in real-space and time are limited. Here, we use magnetic domain walls formed in permalloy artificial square lattice to demonstrate that single-shot Bragg coherent diffraction imaging (BCDI) is a powerful approach for this purpose. Our BCDI reconstructed images revealed milliseconds time-scale dynamics of the antiphase magnetic domain walls traveling through an antiferromagnetically ordered sample.

\end{abstract}

%
\maketitle


Controllable magnetic domain walls are a promising approach toward engineering novel magnetic memory and logic devices due to their low energy cost of operation \cite{zhao2011domain, allwood2005magnetic}. For example, magnetic domain walls can travel with almost zero energy cost, and therefore domain wall-based logic gates can operate with significantly lower power consumption compared to transistor-based bit-flipping. In addition, magnetic domain walls hold promise for cost-effective memory devices because they can be created, manipulated, and erased freely \cite{parkin2008magnetic}. With recent developments in multiferroic materials \cite{valencia2011interface, meier2015functional} whose domain wall motion and dynamics can be manipulated through heat, electric potential, and electric/magnetic field, there is growing promise for their integration into next-generation device architectures.

Visualization of domain walls and their motion provides direct information on the response to driving sources. Studies show that domain wall dynamics can appear in a wide time scale, ranging from nanoseconds to hours \cite{evans2020domains, scott2017superdomain}. Imaging techniques equipped with both magnetic and dynamical sensitivity are needed, but current microscopic methods are limited. Lorentz transmission electron microscopy (LTEM) \cite{phatak2025automated, shimojima2021nano} and photoemission electron microscopy (PEEM) \cite{rivelles2025moving} are powerful real-space probes of magnetic textures, but each has practical limitations. LTEM generally requires electron-transparent samples and is primarily sensitive to magnetic induction, while PEEM is surface sensitive and often limited by photon flux for time-resolved full-field imaging. Although recent LTEM studies have reached sub-millisecond or faster time scales in selected geometries, imaging thermally driven domain-wall motion in extended artificial spin systems remains challenging.

Here, we demonstrate single-shot lensless imaging of magnetic domain wall motion by using Bragg coherent diffraction imaging (BCDI) to an artificial magnetic square lattice with an antiferromagnetic (AF) ground state. Bragg-CDI is an x-ray diffraction technique that leverages the coherence of an x-ray source \cite{robinson2009coherent,pfeifer2006three}. When a coherent x-ray is used, a speckle pattern forms under the envelope of an AF Bragg peak. The speckle patterns are unique to AF domains and domain walls in the illuminated area. From this speckle pattern, we can reconstruct a real-space image of the magnetic domains through a process called phase retrieval. 
Our work is unique in that we employed the resonant property of x-rays and used both structural and magnetic Bragg peaks simultaneously to reconstruct a real-space movie of domain wall motion traveling across the sample. This approach distinguishes our work from previous coherent x-ray studies that primarily tracked magnetic fluctuations or static domain textures without frame-by-frame real-space reconstruction of individual domain-wall propagation.


\section*{}

Artificial spin ice (ASI) systems are two-dimensional arrays of dipolar-coupled nanomagnets designed to simulate the emergent properties of quantum materials such as geometrical frustration \cite{wang2006artificial, nisoli2013colloquium, skjaervo2020advances}. In particular, nanometer-thick square ASI, consisting of nanomagnets arranged on a square lattice (see sketch in Fig.~\ref{fig_scattering}~(a)), exhibits an AF ground state with a magnetic ordering transition occurring close to room temperature \cite{chen2019spontaneous, woods2021switchable, sendetskyi2019continuous}. Antiferromagnets are promising candidates for spintronic applications because they exhibit fast magnetic dynamics, low sensitivity to external magnetic fields, and the absence of stray fields \cite{gomonay2016high}.

\begin{figure}
    \includegraphics[width=0.495\textwidth]{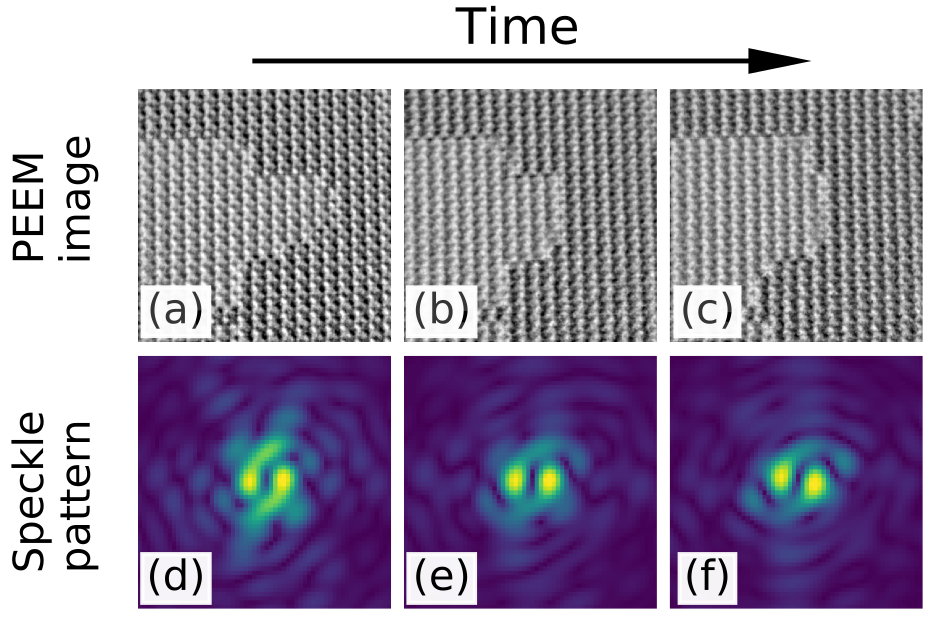}
    \caption{(a)-(c) Time evolution of antiphase domain boundaries observed with PEEM and (d)-(f) corresponding simulated speckle pattern that will appear at AF Bragg peak position in momentum space. The grey overlays on PEEM images are guides to the eye to emphasize two domains that are separated by an antiphase domain boundary (F macrospin-chain). The PEEM images are 15~$\mu$m x 15~$\mu$m each. }
    \label{fig_PEEM}
\end{figure}

While various types of domain walls can form in magnetic materials \cite{meier2015functional}, domain walls in square ASI consist of ferromagnetic (macro)spin chains separating two antiphase magnetic domains, just as in multiferroics \cite{jungwirth2018multiple, wei2013ferromagnetic,venkateswaran2007magnetic,choi2010insulating, hals2011phenomenology,gomonay2016high}. Magnetic domain wall propagation is then a collective spin-flip of the nanomagnets at the domain wall. In Fig.~\ref{fig_PEEM}~(a)-(c), we show a time series of PEEM images, taken 20 seconds apart, displaying thermal fluctuation of an antiphase domain wall in square ASI close to Néel temperature, $T_N$. In the figure, two antiphase domains are distinguished using two overlay colors. Here, we demonstrate that BCDI can image domain-wall motion with higher temporal resolution by first collecting coherent x-ray scattering in momentum space.


\subsection*{Coherent x-ray scattering from artificial structure}

\begin{figure}
    \includegraphics[width=0.45\textwidth]{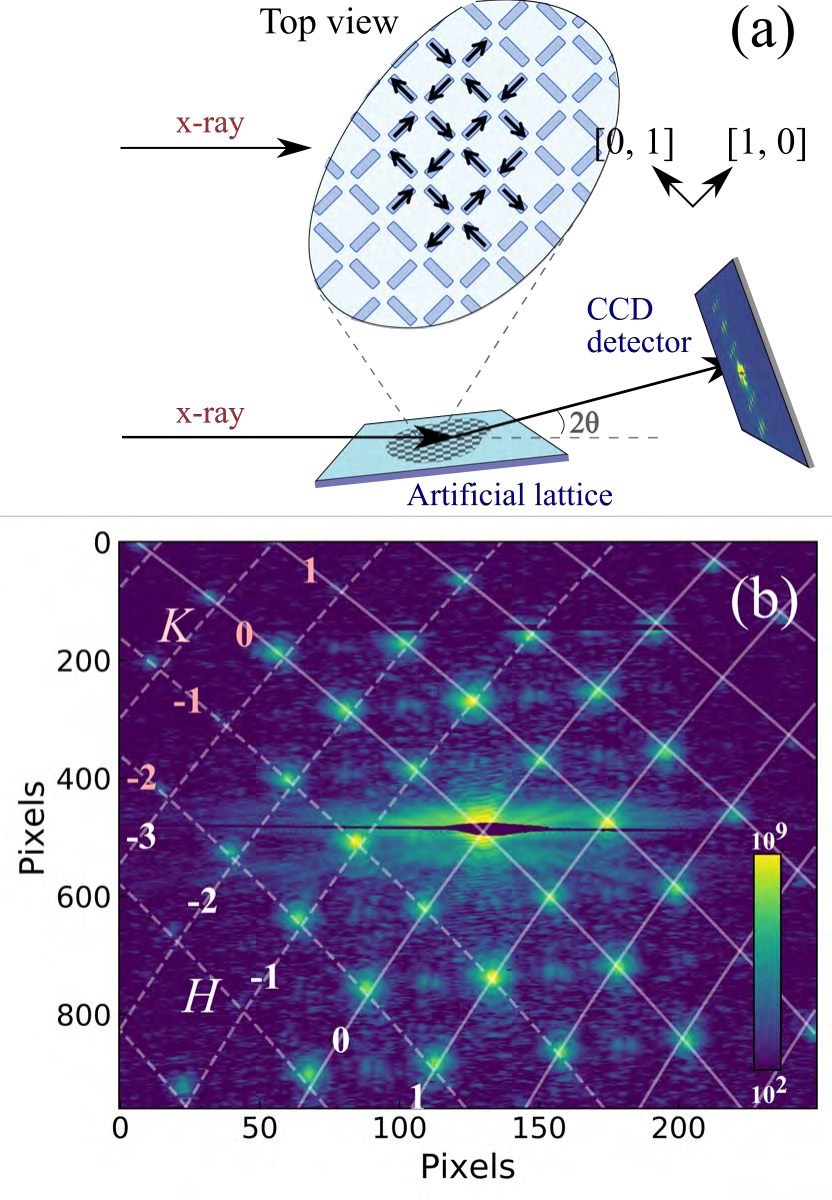}
    \caption{Resonant x-ray scattering with coherent x-ray source. (a) A sketch of the scattering geometry with incident/detector ($2\theta$) angles = 10/20 degrees. The \textit{top view} sketches a small zoomed-up area of the sample to clarify lattice and AF macrospin direction as well as x-ray direction projected to the lattice. (b) Corresponding diffraction pattern that appears on the CCD detector. The r.l.s. coordinates calculated from scattering setups are overlaid on top. Half-integer AF diffraction peaks at $(H/2, K/2)$, and their speckle patterns are clearly visible.
    The AF Bragg peak speckle patterns closely resemble the simulated speckle patterns shown in Fig.~\ref{fig_PEEM}~(d)–(f) (a zoomed-in view is provided in the Supplementary Information).}
    \label{fig_scattering}
\end{figure}

X-ray scattering converts real-space charge and spin information into a signal in momentum space. The scattering rate from an initial to a final state of a system can be described by Fermi’s golden rule, which involves the squared magnitude of the matrix element involved. In this process, the phase information of the sample is lost, and only the amplitude of the transition probability survives. Consequently, real-space images cannot be obtained simply by performing an inverse Fourier transform of a diffraction pattern. Lensless imaging encompasses a class of techniques designed to recover this missing phase information through collective computational algorithms. Modern lensless imaging methods rely on high transverse and longitudinal coherence of the x-rays, which produces high-contrast speckle patterns containing the information needed for accurate phase retrieval.

Coherent x-ray scattering from periodic electronic and magnetic textures results in complex interference (speckle) patterns at their Bragg peaks. The speckle patterns are the squared Fourier transform of real-space textures or domains of interest. As we can see in the simulation in Fig.~\ref{fig_PEEM}~(d)-(f), a small fluctuation of the domain wall gives a clear change in the corresponding speckle patterns. Coherent x-ray scattering has been widely used to probe domain configurations and their fluctuations through the speckle patterns \cite{shpyrko2007direct,klose2022photon, chen2016remarkable,wu2021real,kim2022real, bluschke2022imaging, klose2023coherent}. Bragg CDI is a technique to use speckle diffraction patterns formed in the momentum space and `reconstruct' real-space images by retrieving their 'lost phase'. \cite{fienup1982phase,miao1999extending,chapman2006femtosecond, marchesini2003x, robinson2025spatio}

In BCDI reconstruction, constraints are prior information or physical boundaries imposed during phase retrieval to guide the algorithm toward a unique and physically meaningful solution. A clear sample boundary is one example of a highly effective constraint for the phase retrieval. To create a well-defined sample boundary, we fabricated a square ASI using permalloy on a Si substrate such that the whole ASI pattern can be overfilled by the incident x-ray beam. The experimental concept and the diffraction geometry are illustrated in Fig.~\ref{fig_scattering}~(a). A coherent x-ray beam with 10~$\mu$m diameter overfills the whole fabricated sample at a 10-degree incidence angle. This geometry maximizes resonant magnetic scattering cross-section at Fe $L_3$ absorption edge while suppressing huge x-ray reflection that appears at low incidence angles. A sample image taken by scanning electron microscopy is shown in Fig.~\ref{fig_recon_one}~(a), and the corresponding x-ray diffraction pattern below $T_N$ is displayed in Fig.~\ref{fig_scattering}~(b). The center of the diffraction pattern corresponds to the specular reflection of the incident x-rays, indexed as $(H,K) = (0,0)$. The overlaid $(H,K)$ contour map was calculated from the diffraction geometry and x-ray wavelength. We can clearly see lattice diffraction peaks at integers $(H,K) = (n,m)$ and magnetic diffraction peaks at $(H,K) = (n/2, m/2)$ whose AF lattice doubles structural lattice periodicity.

The resonant x-ray scattering length for magnetic materials can be simplified and written as $f = f_c + iR_1(\bm{k}_o \cdot \bm{m})$, where $f_c$ is the charge scattering from the square ASI, and $R_1(\bm{k}_o \cdot \bm{m})$ is the contribution from magnetic scattering from the magnetic moments in the nanomagnets. $R_1$ is an x-ray energy-dependent matrix that determines the intensity of magnetic scattering contribution. The x-ray intensity is proportional to $|f|^2$, and therefore, in general, charge and magnetic signal interfere. Here, we used $\sigma$-polarization of incident x-rays, in which the charge scattering predominantly remains in the $\sigma \rightarrow \sigma$ channel, whereas the magnetic scattering rotates the polarization into the $\sigma \rightarrow \pi$ channel. The charge and magnetic scattering, therefore, do not interfere and can be decoupled. Hence, the intensities of structural and AF Bragg peaks in Fig.~\ref{fig_scattering}~(b) are simply a linear sum of each other. This allows us to take a two-step process to reconstruct both magnetic domains and the artificial lattice. We first reconstructed the sample shape and lattice using only the speckles at structural Bragg peaks. Then we reconstructed the AF domain textures using the speckles at the AF Bragg peaks. A step-by-step reconstruction is detailed below.

\subsection*{Reconstruction of artificial lattice} \label{sec:recon_lat}

\begin{figure}
    \includegraphics[width=0.45\textwidth]{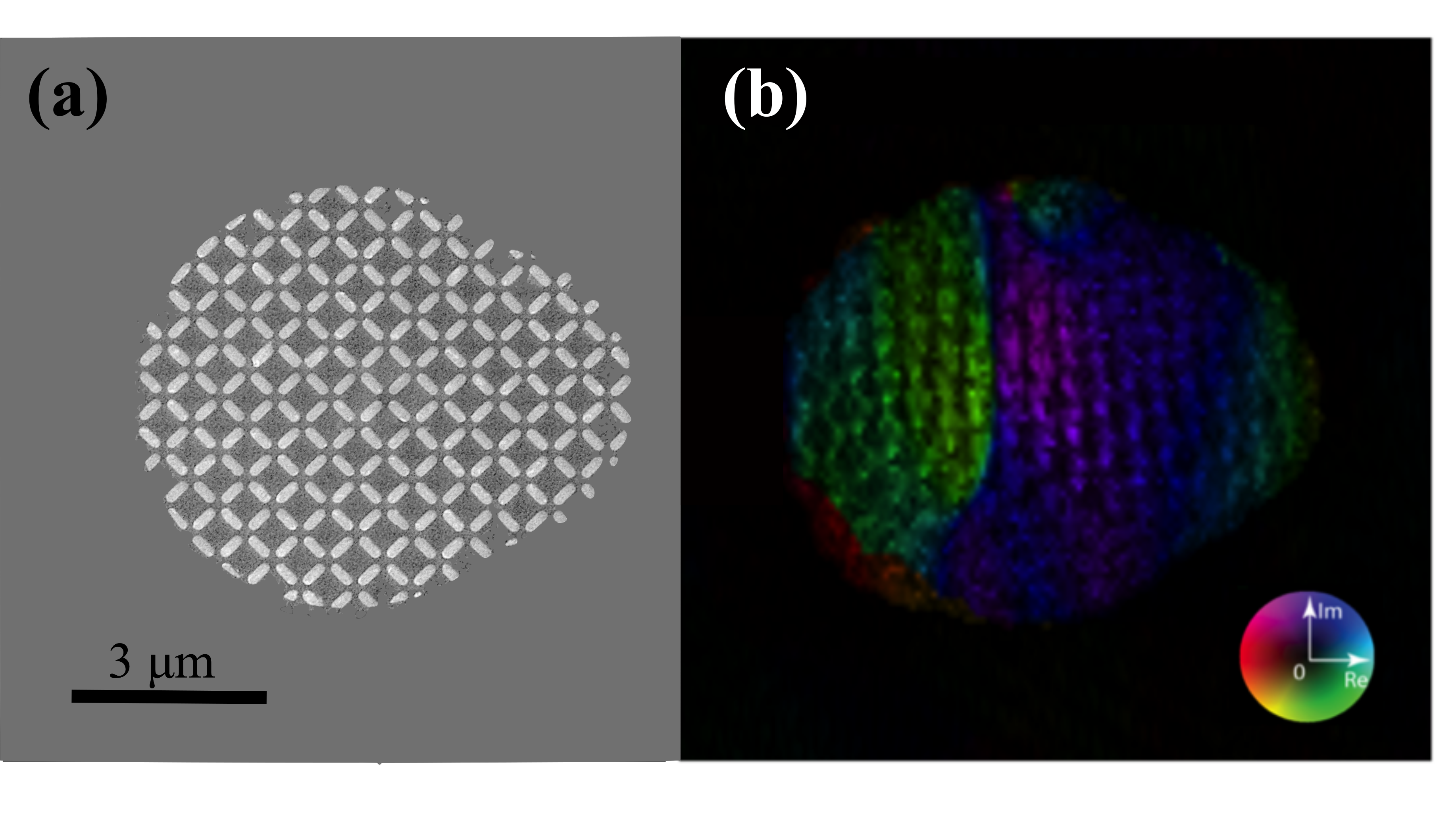}
    \caption{(a) An SEM image of the fabricated square ASI sample and (b) its phase reconstructed image using BCDI. In (b), the reconstructed amplitude (sample shape and the lattice) and phase (magnetic domains) are represented by brightness and color of the image. The wheel-shaped colormap spans the phase of $2\pi$.}
    \label{fig_recon_one}
\end{figure}

Charge scattering -- the small-angle diffuse scattering around the $(H, K) = (0,0)$ along with the structural Bragg peaks -- contains real-space information about the sample shape and the artificial lattice. We used charge scattering by masking out the AF peaks from the diffraction pattern, and reconstructed the sample image. A combination of difference map and error reduction algorithms was employed for iterative phase retrieval \cite{elser2003random,fienup1982phase}. This reconstruction resembles transmission CDI, which has been demonstrated at the CSX beamline at NSLSII (see Supplemental Information). 
The sample SEM image in Fig.~\ref{fig_recon_one}~(a) was used as the initial guess for the sample shape, known as the region of support, or `support'. However, during the reconstruction, we did not fix the support, allowing the sample boundary to evolve into the actual sample shape seen by x-ray illumination.  Without a fixed support, we successfully reconstructed both a clear sample boundary and the structural lattice. Figure~\ref{fig_recon_one} compares the SEM and the reconstructed image of the square ASI sample. In the reconstructed image, the amplitude (brightness) corresponds to the lattice reconstruction. Real crystals in nature can have structural domains or strains that disturb the perfect periodicity of the lattice, resulting in phase differences. In our case, no phase variation is expected because the artificially patterned sample is in a perfectly aligned square lattice. The reconstructed phase from charge scattering is therefore not shown in the figure, as it is constant. The phase (color) in the image corresponds to magnetic domains, which will be discussed in the next section.

\begin{figure*}
    \includegraphics[width=\textwidth]{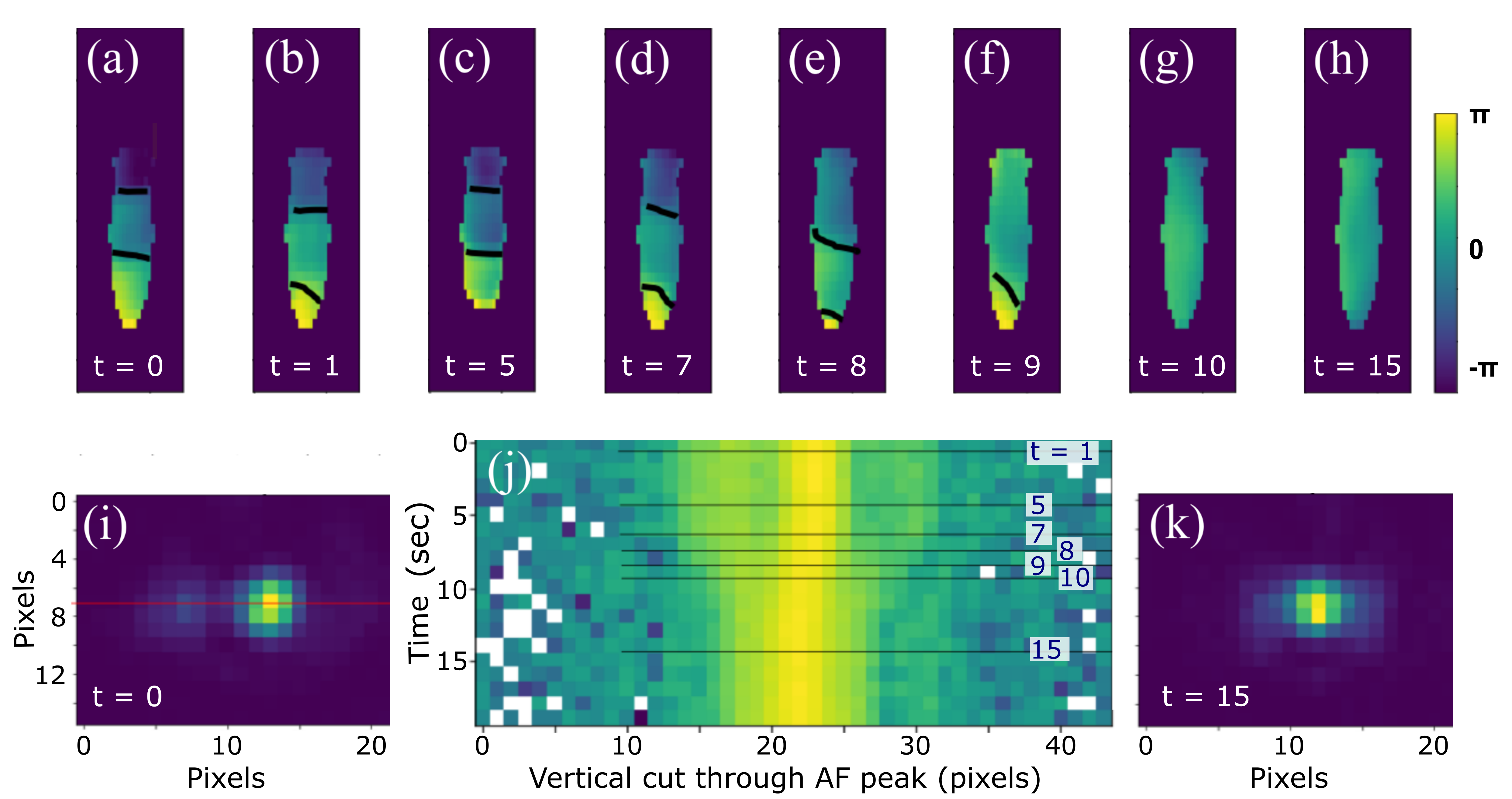}
    \caption{BCDI reconstruction of real-space domain wall motion. (a)-(h) Single-shot BCDI reconstructed images of AF domains at different time slices in the 'waterfall plot' -- time evolution of the speckle patterns over time (j), whose x-axis is a cut through the AF speckle patterns and y-axis is time evolution. (i, k) Detector speckle pattern at the AF peak at times 0 and 15 with an exposure time of 0.1~s. The red line in (i) indicates the cut used in the x-axis of the 'waterfall plot' in (j). Black lines in (a)-(h) are guides to the eyes showing the positions of antiphase domain walls. Times are all in units of seconds, and $t$ is set to $0$ at the first frame of reconstruction.}
    \label{fig_recon_time}
\end{figure*}

\subsection*{Reconstruction of magnetic domains}

Square ASI undergoes a magnetic phase transition from a paramagnetic (PM) phase to an AF phase at $T_N$. The value of $T_N$ depends on the thickness of permalloy, and in our case, $T_N \sim$ 425~K with a thickness of 2.4~nm. Owing to the size and uniform fabrication, our samples naturally form a single-domain ground state when cooled slowly. Antiphase domain walls were introduced by rapidly quenching the sample through $T_N$. The resulting frozen AF domain texture appears as a speckle pattern at the AF Bragg peaks as in Fig.~\ref{fig_scattering}~(b) (see zoomed-in image in Fig.~S1). Notably, the observed speckle pattern closely resembles our simulated speckle pattern generated from PEEM images in Fig.~\ref{fig_PEEM}~(d)-(f).

To reconstruct the image of magnetic domains, we take advantage of knowing the exact shape of the sample boundary obtained in the previous section. Using the reconstructed sample boundary as a fixed support, an image of AF domains was reconstructed using speckles under $\textbf{\textit{q}} = (H, K) = (0.5, -1.5)$ magnetic Bragg peaks. The image reconstructed from a magnetic Bragg peak at vector \textbf{\textit{q}} corresponds to the sample magnetic texture projected onto a plane perpendicular to \textbf{\textit{q}}. In our case, we assume an uniform magnetic texture along the $z$ (out-of-sample) direction due to the thickness of the sample. 

Phase retrieval was performed using the difference map and error reduction algorithms. A random phase and a constant amplitude were assumed for the initial guess because magnetic textures should be uniform except at the antiphase domain walls. Fig.~\ref{fig_recon_one}~(b) shows the reconstructed magnetic domain texture, with the phase color-overlaid on the lattice reconstruction. The reconstructed amplitude was uniform except at domain walls, as expected (not shown). We can identify four magnetic domains in green and purple colors separated by three antiphase domain walls. The color indicates the phase of the magnetic domain, ranging from 0 to 2$\pi$ as indicated with a circle at the bottom. Green and purple colors in the reconstructed image correspond to a $\pi$ shift in phase, and therefore correctly display antiphase magnetic domains as expected.

\subsection*{Domain wall dynamics}

Finally, we present the dynamic properties of the domain walls. The single-shot technique used here allows us to reconstruct magnetic textures for every detector time frame. The detector acquisition frequency, therefore, determines the temporal resolution of dynamics, provided that sufficient speckle intensity is available for reconstruction. Here, we demonstrate that with this technique, we can track domain-wall fluctuation and its travel across the square ASI sample.

In order to visualize domain wall travel, we extended the length of the square ASI sample to allow longer domain wall travel distance. The extended sample (8~$\mu$m x 40~$\mu$m) remained well within the incident x-ray footprint. We tuned the sample temperature close to $T_N$ at $T = 375~K$ to produce sufficient thermal energy to drive spontaneous domain-wall motion. The number of domain walls and their fluctuations increase as the sample temperature approaches $T_N$ \cite{chen2019spontaneous}. To better demonstrate and visualize domain wall movement in both momentum and real space, we carefully chose a temperature such that only a few domain walls are created in the sample. A detector acquisition rate of 0.1~$s$ was used with a maximum photon count of $\sim$60 photons per second per pixel. Fig.~\ref{fig_recon_time}~(i)-(k) present speckle evolution over time. Panels~(i) and (k) show the speckle patterns at the AF Bragg peak at $t = 0~s$ and $t = 15~s$, respectively. The time evolution of the speckles at a line-cut (red line) in (i) is displayed in (j). We can see that the three speckles that initially appeared at $t = 0~s$ gradually merge into a single speckle by the time $t = 10~s$.

Fig.~\ref{fig_recon_time}~(a)-(h) show the time evolution of domain walls obtained from the reconstruction. The image reconstruction for the first detector frame was performed following the same steps as in Fig.~\ref{fig_recon_one}~(b). For the subsequent detector frames at time $t_i$, we used the reconstructed image from the previous time frame $t_{i-1}$ as the starting point for reconstruction and only used the AF Bragg peaks, combined with the error reduction algorithm, to reconstruct domain wall motion. This method is sufficient because the domain wall displacement in subsequent time frames is small. To achieve a clear speckle pattern, we averaged signals from three AF Bragg peaks at the same $|\bm{q}|$ to maximize the signal-to-noise ratio. In the reconstructed images, the domain walls are highlighted with black lines as a guide to the eyes. Only magnetic domains are shown for clear visualization of the domain wall dynamics. In agreement with the waterfall plot, we observe three magnetic domains at $t$ = 0 (a), whose domain walls travel out of the sample by $t$ = 10 (g), resulting in a single-domain sample.

\section*{}

The resolution of the reconstructed image is determined by the extent of the resolvable speckle pattern in momentum space included in the reconstruction. This corresponds to $\sim$ 150~nm resolution for the lattice reconstruction, and $\sim$ 1.2~$\mu$m resolution for magnetic domain reconstruction. We note that 1.2~$\mu$m is not the minimum domain size resolvable with this technique. In our work, we intentionally created a small number of magnetic domains to present a one-to-one comparison of speckle patterns and reconstructed images. The existence of large domains in the sample means a long correlation length, resulting in sharper AF Bragg peaks. The number of speckles at the peak therefore, also decreases accordingly \cite{robinson2020domain}. The reconstruction of magnetic-domain dynamics in Fig.~\ref{fig_recon_time} demonstrates the capability of BCDI to determine the position and propagation of extended antiphase domain walls, rather than to resolve fine internal domain-wall structure. Materials in nature such as multiferroics and other strongly correlated condensed matter systems are known to form short-range correlation \cite{chen2016remarkable}. Such systems studied with soft x-rays can be resolved with spatial resolutions of $\sim$~100~nm, depending on the speckle intensity. 

BCDI revealed that the domain walls can nucleate from the top or the bottom of the sample and travel vertically through the elongated sample. This is explained by the ballistic motion of the domain walls \cite{chen2019spontaneous}. The nanomagnets' magnetization (macrospins) at the edge of the sample needs less energy to flip. Therefore initially, domain walls are nucleated at the edge of the sample by thermal energy. Once the magnetization of the macrospin is flipped at the sample edge, it can either be annihilated or propagate one lattice unit by a collective spin-flip of neighboring macrospins. This will create a domain wall in the sample. This domain wall will propagate ballistically through the sample unless there is a defect to scatter the domain wall. This is why domain walls travel unidirectionally in our reconstruction. This directionality reflects the lower energy cost of flipping fewer collective macrospins perpendicular to the long axis.

\bigbreak

Extending the current work to materials in nature, BCDI can be a powerful tool for visualizing nanoscale structural and magnetic orders. Precise temperature tuning near magnetic transitions provides a way to control domain-wall mobility and explore thermally driven fluctuations systematically. Combining this approach with electric field or laser excitation will enable us to trigger and track domain-wall motion under nonequilibrium conditions in real space. For example, studying domain-wall scattering by defects is particularly relevant in the context of race-track memory devices.

Improving time and spatial resolution is critical for resolving fast motion at small features. Higher coherent flux from 4th generation synchrotrons and faster detectors such as hybrid pixel array detector \cite{chushkin2025achieving} have high potential to increase temporal resolution down to the microsecond scale.  These advances will make BCDI relevant across a wide range of materials, from artificial spin systems to multiferroics and charge density wave compounds. Looking forward, combining BCDI with x-ray free-electron lasers will bring femtosecond time resolution and nanometer sensitivity to magnetic-domain imaging, enabling direct visualization of ultrafast processes relevant to spin transport, memory applications, and phase transitions in quantum materials \cite{yu2024ultrafast}.

\appendix
\section*{Methods}
\subsection*{Sample preparation \label{sample} and characterization}
A square permalloy (Ni$_{0.8}$Fe$_{0.2}$) lattice with 2.4~nm thickness capped with 1.5~nm of Al, was fabricated on a silicon nitride substrate using standard e-beam lithography lift-off techniques. Silicon nitride passes most of the incident x-ray photons and therefore reduces the reflection that gives a large background. The block-spin dimensions were 470~nm long and 170~nm wide, with a lattice constant of $a = b = $ 600~nm. The asymmetric sample shape was introduced to ensure unique reconstruction during phase retrieval. The fabricated samples were inspected using SEM to ensure sample quality. 

\subsection*{Coherent X-ray scattering \label{scattering}}
Coherent x-ray diffraction measurements were performed at 23-ID-1 at the National Synchrotron Light Source II. A 10~$\mu$m pinhole was placed in the beam-path, $\sim$~6~mm from the sample, to establish essentially perfect transverse coherence and longitudinal coherence length of $\sim$~2~$\mu$m at the sample.
The sample was positioned at a $\theta = 10^{\circ}$ glancing angle with respect to the incident $\sigma$-polarized x-ray beam propagation direction to enhance the in-plane x-ray magnetic cross-section with the [0, 1] axis in the scattering plane. The detector was centered on the specularly reflected (zero-order) beam. The sample is fabricated with an elongated shape ($\sim$~8~$\mu$m~x~40~$\mu$m) in order to maximize the scattering volume, but it is still smaller than the x-ray beam footprint so that the x-ray beam profile overfills the sample. This geometry minimizes artifacts from drift in beam or sample positions due to temperature change. The AF Bragg peaks at $(H/2, k/2)$ were measured using resonant enhancement from the magnetic Fe edge at 707~eV. Diffracted photons were collected as a function of temperature using a 1 megapixel fast CCD detector with a readout rate of 10 Hz and 30~$\mu m$ x 30~$\mu m$ pixel size, placed 340~mm from the sample.

\bibliography{refs}

\begin{acknowledgments}

This work was supported by the U.S. Department of Energy, Office of Science, Office of Basic Energy Sciences under Award Number DE-SC0024346. This research used the 23-ID-1 (CSX) beamline of the National Synchrotron Light Source II, a U.S. Department of Energy (DOE) Office of Science User Facility operated for the DOE Office of Science by Brookhaven National Laboratory under Contract No. DE-SC0012704 and resources made available through BNL/LDRD\#19-013. This research used resources (beamline 11.0.1.1, PEEM-3) of the Advanced Light Source, a U.S. DOE Office of Science User Facility under contract no. DE-AC02-05CH11231. Use of the Center for Nanoscale Materials, an Office of Science user facility, was supported by the U.S. Department of Energy, Office of Science, Office of Basic Energy Sciences, under Contract No. DE-AC02-06CH11357. Work at the Molecular Foundry was supported by the Office of Science, Office of Basic Energy Sciences, of the U.S. Department of Energy under Contract No. DE-AC02-05CH11231.  This work was performed in part at the University of Kentucky Center for Nanoscale Science and Engineering, Electron Microscopy Center, and Center for Advanced Materials, members of the National Nanotechnology Coordinated Infrastructure (NNCI), which is supported by the National Science Foundation (ECCS-2025075).
\end{acknowledgments}

\subsection*{Author contributions}

X.M.C. J.S.W, S.R., I.R., L.E.D., and J.T.H. conceived and designed the research. J.S.W, S.D., R.D., D.A.C., W.K.K., and U. W. contributed to sample fabrication and characterization.  X.M.C., J.S.W., A.B., and C.M. contributed to the coherent X-ray scattering experiments.  X.M.C. and R.V.C. contributed to the X-ray photoemission microscopy experiments.  X.M.C. conducted data analysis.  All authors contributed to the manuscript.   

\subsection*{Competing financial interests}
The authors declare no competing financial interests.

\end{document}